%% file: rates.tex
\documentstyle[12pt,aasms4]{article}

\def\Lbsun{\hbox{$\rm\thinspace L_{B\odot}$}}
\def\Mbsun{\hbox{$\rm\thinspace M_{B\odot}$}}
\newcommand{\gapprox}{{_> \atop{^\sim}}} 
\newcommand{\lapprox}{{_< \atop{^\sim}}} 

\slugcomment{To be published in December 10, 1996 issue of  
Astrophysical Journal Vol 473.}
\begin{document}

\title{
The Type Ia Supernova Rate at $z\sim 0.4$}


\author{ 
R. Pain\altaffilmark{1,2},
I. M. Hook\altaffilmark{1,3},
S. Deustua\altaffilmark{1},
S. Gabi\altaffilmark{1},
G. Goldhaber\altaffilmark{1,4},\\
A. Goobar\altaffilmark{1,6},
D. Groom\altaffilmark{1},
A. G. Kim\altaffilmark{1}, 
M. Y. Kim\altaffilmark{1},
J. C. Lee\altaffilmark{1},
C. R. Pennypacker\altaffilmark{1,5}, \\
S. Perlmutter\altaffilmark{1,4},
I. A. Small\altaffilmark{1}, 
R. S. Ellis\altaffilmark{7},
R. G. McMahon\altaffilmark{7}, 
K. Glazebrook\altaffilmark{9},\\
B. J. Boyle\altaffilmark{8},
P. S. Bunclark\altaffilmark{8},
D. Carter\altaffilmark{8}, 
M. J. Irwin\altaffilmark{8}\\
(Supernova Cosmology Project)
%
}
\altaffiltext{1}{E. O. Lawrence Berkeley National Laboratory, 
Berkeley, California 94720}
\altaffiltext{2}{LPNHE, CNRS-IN2P3 and Universit\'es Paris VI \& VII, France}
\altaffiltext{3}{Astronomy Dept, UC Berkeley CA94720}
\altaffiltext{4}{Center for Particle Astrophysics, U.C. Berkeley, California 94720}
\altaffiltext{5}{Space Sciences Laboratory, U.C. Berkeley, California 94720}
\altaffiltext{6}{University of Stockholm, Sweden}
\altaffiltext{7}{Institute of Astronomy, Cambridge, CB3 OHA, United Kingdom}
\altaffiltext{8}{Royal Greenwich Observatory, Cambridge, CB3 OEZ, United Kingdom}
\altaffiltext{9}{Anglo-Australian Observatory, P.O. Box 296, 
Epping, NSW 2121 Australia}


\begin{abstract} We present the first measurement of the rate of Type
Ia supernovae at high redshift. The result is derived using a large
subset of data from the Supernova Cosmology Project. Three supernovae
were discovered in a surveyed area of 1.7 square degrees. The survey
spanned a $\sim 3$ week baseline and used images with $3\sigma$ limiting
magnitude of $R\sim 23$. We present our methods for estimating
the numbers of galaxies and the number of solar luminosities to which
the survey is sensitive, and the supernova detection efficiency which
is used to determine the control time, the effective time for which the
survey is sensitive to a Type Ia event. We derive a rest-frame Type
Ia supernova rate at $z\sim0.4$ of $0.82\ {^{+0.54}_{-0.37}}\
{^{+0.37}_{-0.25}} $ $h^2$ SNu (1 SNu = 1 SN per century per
$10^{10}$\Lbsun), where the first uncertainty is statistical and the
second includes systematic effects. For the purposes of observers, we
also determine the rate of SNe, per sky area surveyed, to be $ 34.4\
{^{+23.9}_{-16.2}}$ SNe\ $\rm year^{-1} deg^{-2}$ for SN magnitudes in
the range $21.3 < R < 22.3$.
\end{abstract}


%
%

\section{Introduction}


Type Ia supernovae (SNe) may provide one of the best testable distance
indicators at high redshifts, where few reliable distance indicators
are available to study the cosmological parameters.  A direct
measurement of SN rates is therefore important in developing
systematic programs to find and study such high-redshift SN Ia
distance indicators.  Supernova rates at high redshift are also
important for understanding galaxy evolution, star formation rates,
and nucleosynthesis rates.  The dependence of the Type Ia SN rate on
redshift can be used to constrain models for the progenitors of the SN
explosion. For example Ruiz-Lapuente et al (1995) discuss the
correlation of Type Ia rate and explosion time with parent population
age and galaxy redshift.

Beginning with the discovery of SN 1992bi (Perlmutter et al 1995a), we
have developed search techniques and rapid analysis methods that allow
systematic discovery and follow up of ``batches'' of high-redshift
supernovae.  At the time of this analysis, the search had discovered 7
SNe at redshifts $z=0.3$ to $0.5$ (Perlmutter et al 1994, 1995a,
1995b). We report here our first estimates of the SNe Ia rate at high
$z$ based on a subset of this data set. We are currently following 
$>18$ further SNe in the range $0.39<z<0.65$ but
the data collection and analysis of these SNe are not yet complete.

The observing strategy developed for our search compares large numbers
of galaxies in each of $\sim$50 fields observed twice with a
separation of $\sim$3 weeks, thus almost all our SNe are discovered
before maximum light.  This search schedule makes it possible to
precisely calculate the ``control time,'' the effective time during
which the survey is sensitive to a Type Ia event.  On the other hand,
since hundreds of anonymous high-redshift galaxies are observed in
each image, it is more difficult than for nearby SN searches to
estimate the number, morphological type and luminosity of galaxies
searched in a given redshift range.

The method used to calculate the rate can be divided into two main
parts: (i) estimation of the SN detection efficiency and hence the
control time, and (ii) estimation of the number of galaxies and the
total stellar luminosity (measured in $10^{10}$\Lbsun) to which the
survey is sensitive. We have studied our detection efficiency as a
function of magnitude and supernova-to-host-galaxy surface brightness
ratio using monte-carlo techniques.  By comparing galaxy counts per
apparent magnitude interval in our images to the study of Lilly et al
(1995) we have estimated the number of galaxies in a given interval of
redshift and apparent magnitude.  The galaxy counts and efficiency
studies, together with the number of confirmed SN detections in this
set of images yields an estimate of the SN Ia rate at $z\sim 0.4$.


In Section 2 of this paper, we describe the data we have used. Section
3 deals with the determination of the efficiency of the search, and
hence the control times.  Section 4 covers the method for estimating
galaxy counts. In section 5, we derive the Type Ia rate at $z\sim
0.4$; in Section 6, we estimate the systematic uncertainty; and in
Section 7 we discuss the results.

\section{The Data Set}

For this analysis, we have studied a set of 52 similar search fields
observed in December 1993 and January 1994. This is the first sizeable
data set to arise from the Supernovae Cosmology Project. These images
are suitable for a determination of the SN rate since they were obtained
under similar conditions with a single camera at one telescope, and
therefore form a well-defined, homogeneous set.

The data were obtained using the ``thick'' $1242\times 1152$ EEV5
camera at the 2.5m Isaac Newton Telescope, La Palma.  The projected
pixel size is $0.56''$, giving an image area of approximately
$11'\times 11'$. Exposure times were 600s in the Mould $R$ filter, and
the individual images typically reach a $3\sigma$ limit of at least
$R\sim23$mag. Seeing was typically around $1.4''$. The fields lie in
the range $2^h<\alpha(B1950)<14^h$, $\delta(B1950) > -10 ^{\circ}$,
excluding the galactic plane ($|b|\gapprox 30^{\circ}$). Many of the
fields were selected due to the presence of a high-redshift cluster
($z\sim 0.4$).  Suitable clusters and their redshifts were taken from
Gunn, Hoessel \& Oke (1986). The effect of the presence of clusters in
the survey fields is taken into account in the calculation of the SN
rate (see Section 4).

For most fields, two first-look ``reference'' images were obtained,
here called ($ref_1$, $ref_2$), and for all fields two second look
``search'' images ($search_1$, $search_2$) were obtained $2-3$ weeks
after the reference images.  The useful area of this dataset is
defined by the overlap area of the original set of reference images
with the second set of search images. The total useful area covered in
this study is 1.73 sq deg.

The analysis procedure and method for finding SNe can be summarized as
follows.  The images were flat-fielded and zero-points for the images
were estimated by comparison with E (red) magnitudes of stars in the
APM (Automated plate measuring facility, Cambridge, UK) POSSI catalog
(McMahon \&  Irwin 1992).


For the final analysis of the SN light curves for the determination of
$q_0$, the fields containing SNe are re-observed and calibrated using
Landolt standards (Landolt, 1992). However, this calibration is not
available for all our search fields that do not contain SNe, hence the
use of APM calibration for this study.  A comparison of APM E
magnitudes with CCD $R$ magnitudes shows that $E-R$ has a mean of
$-0.2$~mag and rms $0.2$~mag. We therefore applied a 0.2 mag shift to
the APM magnitudes.  The uncertainty in the rate introduced by the
uncertainty in the zero-points is discussed in Section 6.

The search images were combined (after convolution to match the seeing
of the worst of the four images) and the combined reference images
were subtracted from this, after scaling in intensity.  The resulting
difference image for each field was searched for SN candidates. The
main selection criteria was that the object must be a $4\sigma$
detection on the difference image. The candidate list was filtered by
requiring that the object must not move by more than 2 pixels between
the two search images (to remove asteroids), and that the object be a
$2.5\sigma$ detection on the separate difference images (i.e.
$search_1-combined\ ref$ and $search_2- combined\ ref$).  There was no
requirement that the candidate be on a visible host galaxy.  The
remaining candidate SNe on all the images were inspected visually for
obvious problems such as very bright stars nearby, bad columns etc.,
which could affect the photometry. Follow-up photometry and
spectroscopy was used to determine the SN type.

In this subset of the search data three SNe were found, with redshifts
0.375, 0.420 and 0.354. Their properties are summarized in
Table~\ref{sntab}.  SN1994H was discovered in a field centered on a
cluster (Abell 370). Its host galaxy is at the cluster redshift
($z_{clus}=0.373$) at a projected distance of 1.1 arcmin. The
redshifts were determined from spectra of the host galaxies. For the
purposes of this analysis, we assume these are all Type Ia SNe. This
is a likely scenario since Type Ia's are the brightest of the SN
types, and therefore the most likely to be detected at large
distances. The measured light-curves of the three SNe
follow the standard Type Ia light-curve template (Leibundgut,
1988). The data, however, could be consistent with unusually
bright Type IIL supernovae. 
Using our previous estimate of one very bright Type IIL SN for 
ten Type Ia SNe (Perlmutter et al. 1995), we expect only
at most 0.3 Type IIL SNe in the sample. This in agreement with 
our recent discovery of 11 supernovae where 
ten are spectroscopically consistent with being Type Ia SNe and one
with being a type II (Perlmutter et al. 1995c).
For SN1994F we also have a spectrum and it is
consistent with that of a Type Ia SN at the date of observation.
Other spectra we have obtained for the larger sample of SNe discovered in 1994
are also
consistent with those of Type Ia SNe. In addition, in this larger sample, 
one of the SNe for
which no spectrum was obtained was in an elliptical host galaxy, a
strong indicator of Type Ia identification.  A full discussion of the
light curves and spectra will be given in a future paper (Perlmutter
et al, in preparation).

During re-analysis of the data for the purposes of calculating the
rate, another faint ($R=22.5$) object was found. This
candidate SN which was very near detection threshold, had not been 
classified as a SN at the 
scanning stage (final visual inspection) during the first analysis.
The
object shows a fairly large motion between the two search images (0.5
arcsec, $\sim 1$ pixel) which indicates that it is most likely to be a faint
asteroid. Since it was not followed, it 
cannot be ruled out   
as a possible supernova. 
Although unlikely, the possibility that the object is a Type
Ia SN was taken into account in the systematic uncertainty as
discussed in Section 6.  

\section{Control Times and Detection Efficiencies}

A na\"{\i}ve estimate of the control time $\Delta T$ is given by the time
during which the supernova light-curve is above a given threshold
corresponding to the limiting magnitude of the observations. In our
case, this significantly overestimates the control time for the
following reasons.

The data presented here were obtained with an observing strategy
designed to measure $q_0$ by conducting a search for SNe on the rise
(before their maximum light) using a subtraction technique. The signal
on the search image must therefore be significantly larger than on the
reference image, reducing $\Delta T$ from the na\"{\i}ve estimate by
approximately a factor of two.  In addition, the detection efficiency
depends on the host galaxy magnitude, the image quality, the search
technique and strongly on the magnitude of the supernova at the
detection time.

In this analysis we compute a control time equal to the weighted sum
of days during which the SN can be detected, where the weighting is by
the corresponding detection efficiency, $\epsilon$. The control time
is given by $\Delta T=\int \epsilon(t) dt$, where $\epsilon$ is a
function of the observed magnitude $m_{obs}$, which is itself a
function of time $t$ relative to maximum, and $\delta t$, the time
separation of the search and reference images. 
To account for the $z$ distribution of galaxies in the fields, control
times were calculated in bins of $z$ (and host galaxy magnitude).      
We assumed that SN magnitude as a function of time
follows the average of the best-fit, time-dilated and $K$-corrected
Leibundgut Type Ia template light curves. The generalized $K$
correction described by Kim et al. (1996) was used here.

The three light curves we use as example high-redshift SNe have been
calibrated using Landolt standards (Landolt, 1992).  Since these are observed
light-curves, in {\it apparent} magnitudes, no explicit dependence of
our rate result on $H_0$ or $q_0$ is introduced at this stage.  The
control time was computed taking galactic extinction into account for
each field separately. The reddening value for each field $E(B-V)$
were supplied by D. Burstein, (private communication, derived from the
analysis of Burstein \& Heiles, 1982) and were applied to the data
assuming $R_V=3.1$, and $A_R/A_V=0.751$ (Cardelli et al, 1989).


The detection efficiency $\epsilon$ is a complicated function of many
parameters. The efficiency as a function of the SN magnitude depends
on the quality of the subtracted images (seeing, transmission)
together with the detailed technique (convolution, selection criteria)
used to extract the ``signal'' (SNe candidates) from the ``background''
(cosmic rays, asteroids, bad subtractions, etc). In addition, there is
a slight dependence on the host galaxy magnitude.  The detection
efficiency was calculated using a Monte-Carlo method. A synthetic
image was created for every field by adding simulated supernovae to
the search images. The reference images were subtracted from the
synthetic search images using exactly the same software as used for the
supernova search, described in Section 2, and the number of simulated
SNe that satisfy the selection criteria was determined. This technique
allows us to measure detection efficiencies as a function of
supernova magnitude individually for every field, thus taking into
account the other parameters mentioned above. The efficiency derived
in this way includes the effects of parts of the image being unusable
for the SN search, e.g. due to bright foreground stars.

One hundred simulated SNe were placed on each search image, with a
range of SN apparent magnitude, host galaxy apparent magnitude and
locations with respect to host galaxies.  Each simulated SN was
generated by scaling down and shifting a bright star from that image,
with signal-to-noise ratio greater than 50, from the image being
studied.  The position relative to the host galaxy was chosen at
random from a normal distribution with $\sigma$ equal to the half
width at half maximum of the galaxy. The shift of the scaled bright
star to the host galaxy was by an integral number of pixels to
maintain the pixelized point spread function.  Since noise
fluctuations in the sky background strongly dominates the SN photon
noise, it was not necessary to add extra Poisson noise to these
simulated SNe.

\begin{figure}
\plotone{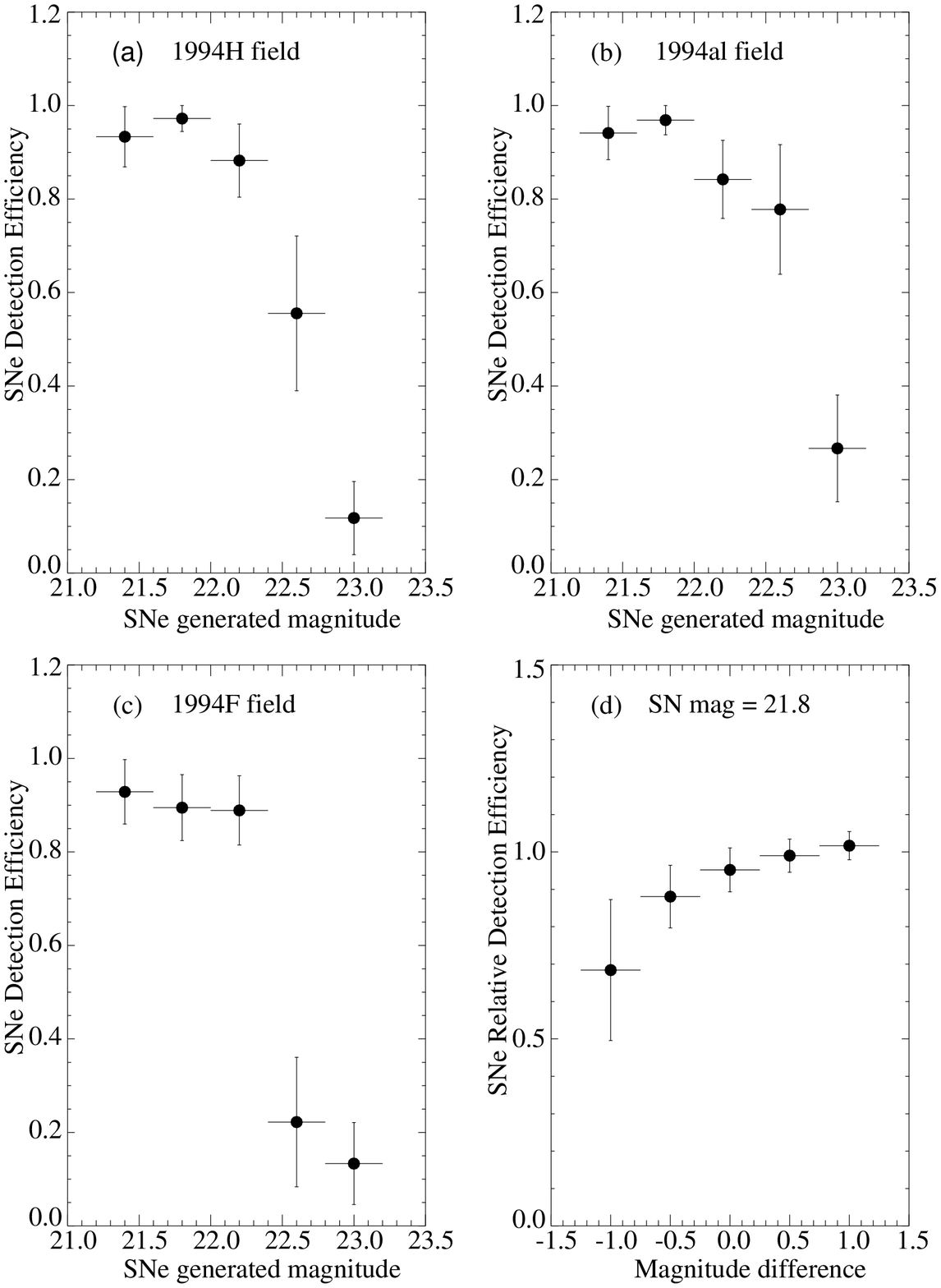}
\caption{(a)-(c) Detection efficiency as a function of magnitude 
for the three difference images in which SNe were found. The
vertical error bars show the $1\sigma$ statistical uncertainty, and the
horizontal bars show the bin ranges (d) Detection efficiency as a 
function of relative SN to host surface brightness. \label{eff}}
\end{figure}

Figure~\ref{eff}(a)-(c) shows the fractional number of simulated SNe
recovered, as a function of SN magnitude (at detection) for the three
fields in which SNe were found. Figure~\ref{eff}(d) shows the
efficiency as a function of relative surface brightnesses of the SN
and host galaxy.  This last parameter gives an indication of the
effect of SN location with respect to the host galaxy. Although this
is a small effect, it was taken into account. For a typical field the
detection efficiency is over 85\% for any added fake stellar object
brighter than $R=22.0$ magnitude (Note that the more recent searches 
of this project have worked with significantly deeper images). 

At this stage we are able to determine the ``survey rate'' of SN
discoveries that a search for Ia SNe can expect to obtain, per square
degree. We give the rate in a range of 1 mag in $R$, centered on the
mean peak $R$ magnitude of the 3 SNe found in this search, $R=21.8$. The
survey rate is given by
 $$ survey\ rate\ (21.3<R<22.3) = {N_{SN}\over{\sum_i area_i \times
\Delta T_i}}.$$ where $N_{SN}=3$ is the number of SNe we found in the
1 magnitude range, and $\Delta T_i$ is the control time for field $i$,
computed for a SN with magnitude $R=21.8$ at maximum.  For example a
value of $\Delta T_i= 21$ days was found for the field containing
SN1994H observed on 1993 December 19 and 1994 January 12, 24 days
apart.

We measure a survey rate for $21.3 < R < 22.3$ of $ 34.4\ {
^{+23.9}_{-16.2}}$ SNe\ $\rm year^{-1} deg^{-2}$ (the error quoted is
statistical only). In practice this translates to 1.73 SNe per square
degree discovered with a 3 week baseline, in data with limiting
magnitude (3$\sigma$) $R\sim 23$.  The total number of galaxies with
$R<23.8$ surveyed in the 52 fields is approximately 32,000.

\section{Galaxy Counts}

In order to compare the distant SN rate with local equivalents, we
need to know the redshifts of the galaxies we have surveyed. We
estimate these in a statistical manner using various groups' analyses
of galaxy evolution.  In this work we use the galaxy counts derived
from the analysis of Lilly et al (1995) to estimate the number of
galaxies sampled as a function of redshift.  We have also carried out
the analysis using the galaxy evolution model of Gronwall \& Koo (eg
Gronwall \& Koo 1995) and that used by Glazebrook et al (1995), to
give an estimate of the sensitivity of our results to the assumed form
of galaxy evolution.

$R$ band counts as a function of redshift were calculated by S. Lilly
based on the analysis of magnitude--redshift data obtained in the
Canada-France Redshift Survey (Lilly et al, 1995 and references
therein). The survey contains $\sim 730$ galaxies with
$17.5<I<22.5$. Lilly et al estimate the expected distribution of
galaxies with redshift and $R$ band magnitude $N(z,R)$, by extrapolation
from the $I$ band data, with the implicit assumption that the galaxy
population does not evolve at redshifts outside the limits of the CFRS
sample. $q_0=0.5$ was assumed for these calculations (the effect of
this assumption on the derived SN rate is discussed in Section 6).
Since the I band is close to the $R$ band, and the magnitude range of
the CFRS sample is comparable to that of our data, this extrapolation
is small.

\begin{figure}
\plotone{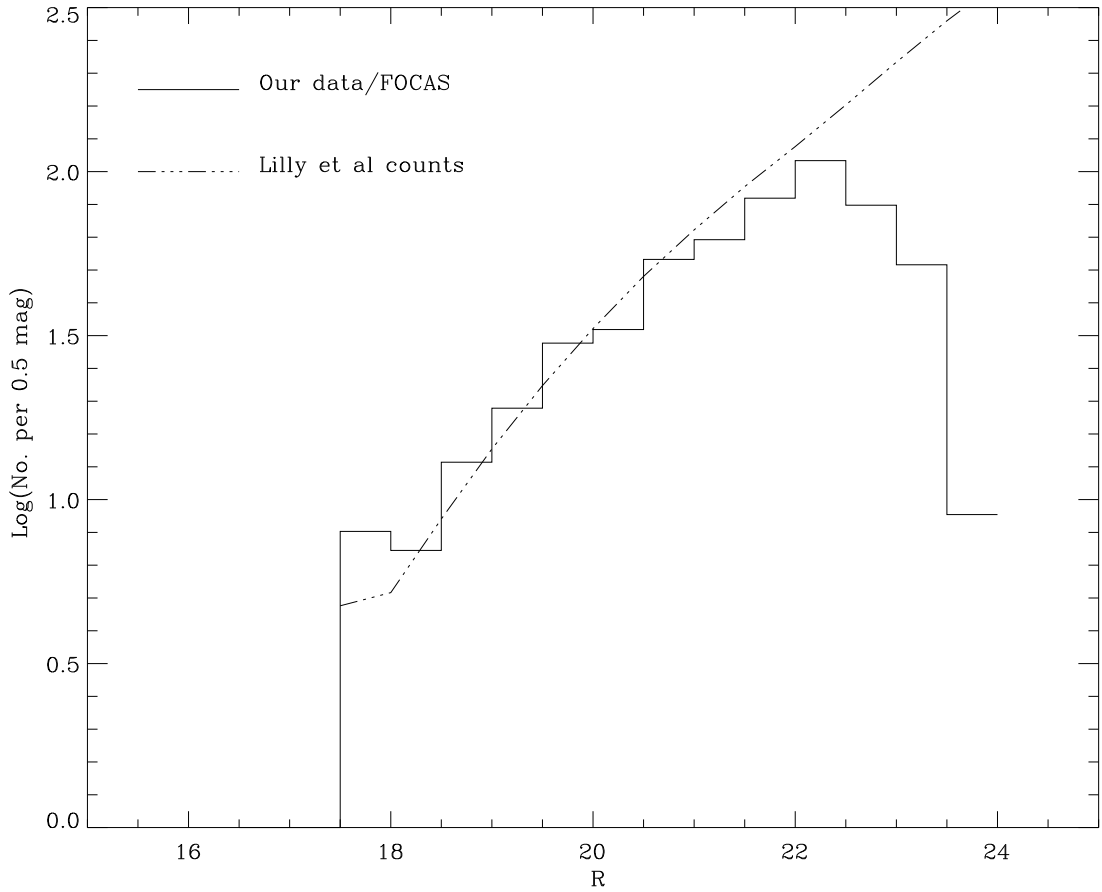}
\caption{Number of galaxies as a function of
magnitude determined from one of our non-cluster images using FOCAS.
The dashed-dot line shows the counts derived from the analysis of Lilly
et al (1995), integrated over the redshift range $0<z<2$, and normalized to
the image area of 0.03 sq deg. \label{nofm}}
\end{figure}

To check that the assumed distribution of galaxies with $R$ magnitude
and redshift, $N(z,R)$, give reasonable galaxy counts compared to our
data, we have plotted the number of field galaxies classified by the
FOCAS software package, as a function of apparent magnitude, on one of
the search images that was {\em not} targeted at a galaxy cluster.
The $R$-band galaxy counts given by the analysis of Lilly et al (1995)
integrated over the redshift range
$0<z<2$ (dash-dotted line) are shown on the same scale, assuming an effective
area for this image of 0.03 sq deg (Figure~\ref{nofm}).
 
\begin{figure}
\plotone{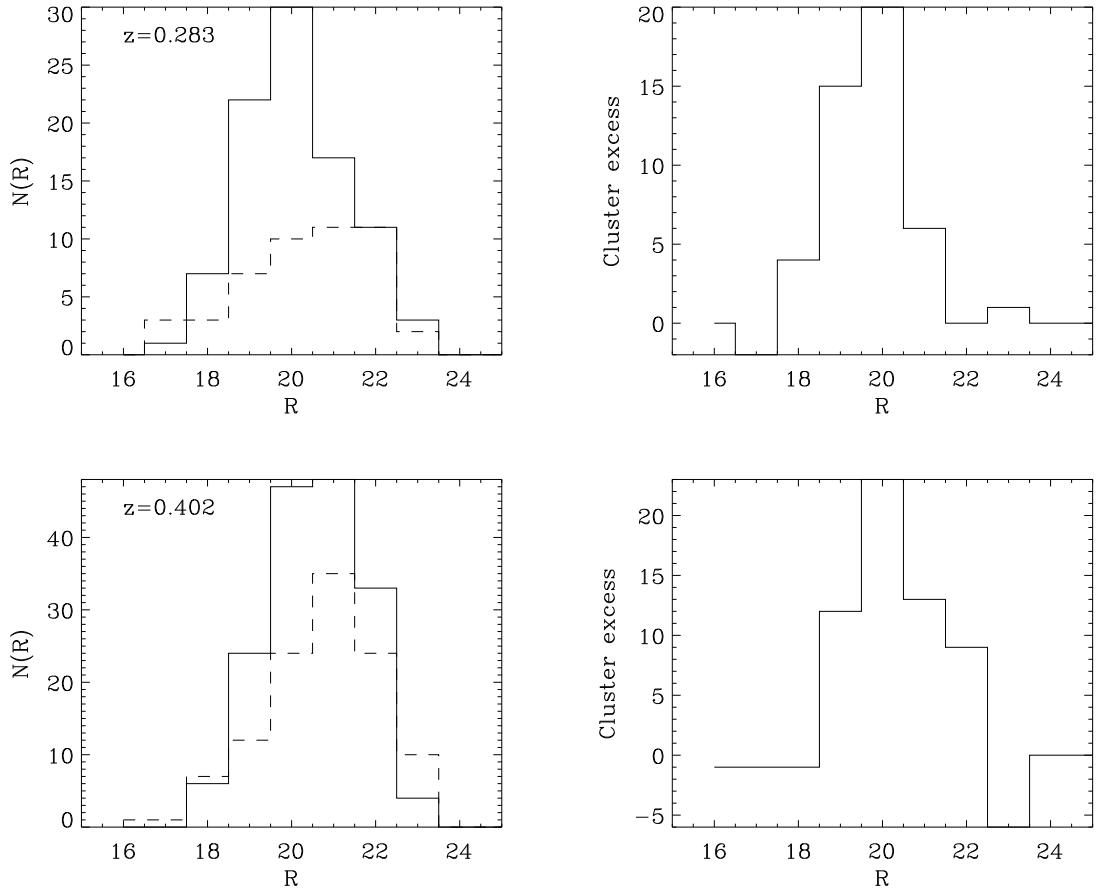}
\caption{Galaxy $N(R)$ in a $500\times 500$ pixel
box containing the cluster (solid line) and $N(R)$ in a similar box away
from the cluster (dashed line) for two fields. The excess cluster counts
are shown in the right hand panels.  \label{cluster} }
\end{figure}

Many of our search fields were chosen specifically to target
high-redshift clusters.  For each of these fields, we estimate the
number of cluster galaxies by counting galaxies as a function of $R$
magnitude in a box of size $500\times 500$ pixels centered
approximately on the center of the cluster as estimated by eye from
the images. The counts in a similar box in a region of the image away
from the cluster were subtracted from the cluster counts to give the
cluster excess counts as a function of $R$ mag. Examples of these
distributions are shown in figure~\ref{cluster}.  Typically a cluster
contributes 10\% of the galaxy counts on an image.  We assign these
galaxies to the cluster redshift, and add the cluster contribution to
the $N(z,R)$ for that image given by the models.

\section{SN Ia Rates}

To compare our derived SN rate with local rates, we express the rate
in units of SNu, the number of SNe per century per $10^{10}$ solar
luminosities in the rest-frame $B$ band. To calculate the rate we
derive the expected redshift distribution of SNe, $N_{exp}(z)$, which
is proportional to the observed SN rate, $r_{SN}(1+z)^{-1}$, where
$r_{SN}$ is the rate in the rest-frame of the supernovae. The expected
distribution is given by $$N_{exp}(z)={r_{SN}\over{1+z}}\sum_R\sum_i
N_{gal}(z,R)_i \times L_B (z,R) \times \Delta T_i (z,R) $$ where $i$
runs over all fields, $R$ is the galaxy apparent $R$ magnitude and
$L_B$ is the galaxy rest-frame $B$ band luminosity in units of
$10^{10}$\Lbsun.  We then fit the observed redshift distribution to
$N_{exp}$ and hence derive $r_{SN}$. Here it is assumed that the
rest-frame rate $r_{SN}$ is constant in the redshift range of interest
($0.3\lapprox z\lapprox 0.5$).  The control times $\Delta T$, in units
of centuries, have been calculated for each field in bins of $z$ and
$R$ (the size of the bins used is 0.5 mag in $R$, 0.05 in $z$). The
derived rate corresponds to a mean redshift given by $$<z> = \int{ z
N_{exp}(z) dz}/\int{N_{exp}(z) dz} $$

To compute the rest-frame $B$ band galaxy luminosities from apparent $R$
magnitudes, we used $B-R$ colors and $B$-band $K$ corrections (which
include the effects of evolution) supplied by Gronwall \& Koo (private
communication). These are based on the models of Gronwall \& Koo
(1995), which give the relative proportions in each bin of $z$ and
apparent $R$ magnitude of three different color classes of galaxies
(defined as `red' $B-V>0.85$, `green' $0.6<B-V<0.85$ and `blue'
$B-V<0.6$).  Note that the combined color, $K$ and evolution correction
is small in the redshift range of interest ($0.3-0.5$) mostly because the
observed $R$ band is close to the rest-frame B-band.  
The appropriate correction for each color class was applied in the
proportions given by the model, and the total luminosity of galaxies
in that bin was computed. In this calculation $\Mbsun=5.48$ and
$q_0=0.5$ were assumed. Table~\ref{ntot} gives the total luminosity in
bins of $z$ and $R$ mag.

\begin{figure}
\plotone{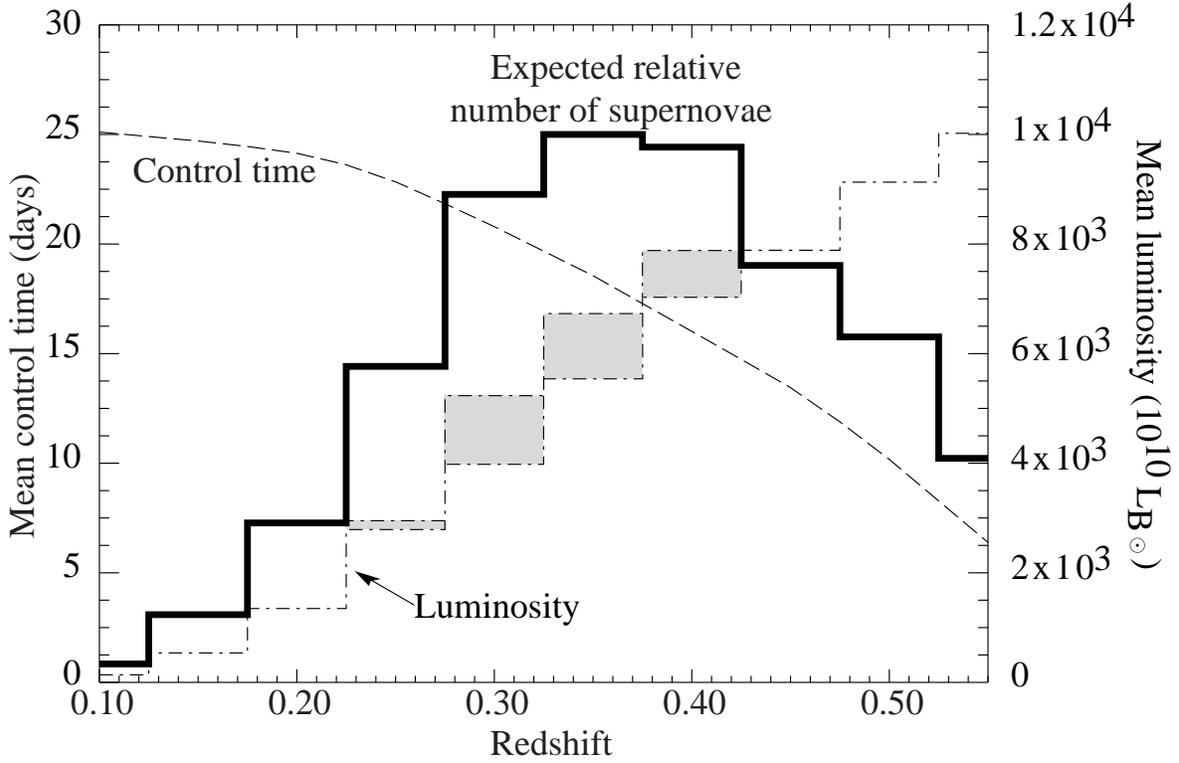}
\caption{The expected number of supernovae as a function
of $z$ (solid
histogram) together with the overall detection efficiency given as a
``control time'' (dashed curve) and the luminosity-weighted number of
galaxies (dash-dot histogram). The contribution to the luminosity from
clusters is shown by the shaded area. The December 1993-January 1994
search was most likely to find SNe with redshifts between $z=0.3$ and
$z=0.4$. Between $z=0.3$ and $z=0.55$, the search was more than 50\%
efficient. Note that our more recent searches go deeper than this
data.  \label{exp} }
\end{figure}

Figure~\ref{exp} shows the expected redshift distribution of SNe,
$N_{exp}(z)$, as calculated above. The detection efficiency as a
function of $z$, expressed as the mean control time $\Delta T$,
averaged over all fields is also shown, as well as the mean galaxy counts
weighted by their B- band luminosities. These two mean quantities are
shown merely for illustration; they are not used in the calculation of
the expected distribution since each field is treated separately and
the results combined.

The rest frame supernovae rate $r_{SN}$ at $z\sim0.4$ was obtained by
fitting the redshift distribution of observed SNe to the expected
distribution, $N_{exp}(z)$, using a maximum-likelihood fit with Poisson
statistics.  The mean redshift corresponding to this rate is
$<z>=0.38$. We derive a value for the SN rate of
$$r_{SN}(z=0.38)=0.82\ ^{+0.54}_{-0.37}\ h^2\ {\rm SNu}$$
where the error is statistical only.

\section{Systematic Uncertainties}
 
Because of the small number of SNe in this first sample, the total
uncertainty in this measurement is dominated by statistics. We have,
nevertheless, estimated the following systematic uncertainties. The
sources studied are listed below, and Table~\ref{systab} summarizes
their contributions.

\paragraph*{Total Luminosity estimate} The total solar luminosity to which
the survey is sensitive was estimated using counts for $N(z,R)$,  which
have statistical uncertainty due to the finite number of galaxies used
in the analysis (e.g. $\sim 700$ in the analysis of Lilly et al, 1995).
The statistical uncertainty in the model contributes about $\pm 0.02 h^2$ SNu
uncertainty in the rate. 

In addition, the luminosity estimation depends on the deceleration
parameter, since $q_0$ enters in the galaxy $N(z,R)$ model/predicted
counts, the $K$ corrections of Gronwall \& Koo (1995) and the luminosity
distance used to calculate the galaxy absolute luminosities. To
estimate the sensitivity of our result to $q_0$, we repeated the
analysis using versions of the Lilly et al counts and the Gronwall \&
Koo model which were calculated for $q_0=0.0$. This value of $q_0$ was
also used to calculate the luminosity distances. The deceleration
parameter does not affect the calculation of control times because the
observed light curves are used in this calculation.  The total effect
is small, and is dominated by the effect on the luminosity
distance. Using $q_0=0.0$ rather than $0.5$ lowers the derived rate by
$ 0.08 h^2$ SNu. Similarly increasing $q_0$ by $0.5$ raises the rate
by a comparable amount.  


The combined uncertainty in the rate due to the luminosity estimate is
therefore $\pm 0.09 h^2$ SNu

\paragraph*{Contribution from clusters} Many of our fields contain a
known cluster in the redshift range $0.3-0.5$.  Four of our fields
contain visible clusters which do not have known redshifts. To
estimate the effect this uncertainty has on the derived rate, we
assigned all the unknown redshifts to $z=0.1$ and in separate analyses
assigned their redshifts 
to $z=0.7$ and $z=0.4$ where this search is most sensitive. The effect
of changing the assumed $z$ from 0.1 to 0.4 is to decrease the rate by
$0.01 h^2 $ SNu. Similarly changing the assumed redshifts from $z=0.4$
to $z=0.7$ lowers the rate by $0.01 h^2$SNu. 
There is also some uncertainty due to the faint
cluster galaxies which are not detected on our images, but which could
host a detectable SN.  We estimate less than a 10\% uncertainty in
calculating the overall contribution to the galaxy counts from these
clusters, giving a contribution of $\pm 0.02 h^2$ SNu to the
uncertainty in SNe rate.

\paragraph*{Extinction Correction} The uncertainty from correcting for
extinction was calculated following the estimate from Burstein \&
Heiles (1982) of the uncertainties in deriving the Galaxy
reddening. The effect on the rate is small and amounts to $\pm 0.01
h^2$ SNu.

\paragraph*{ APM calibration} Although we used measured SN light
curves, calibrated with Landolt standards, to calculate the control
times, the galaxies were calibrated using the less accurate APM
calibration.  Errors in the APM calibration of the fields would thus
alter the determination of the efficiency as a function of
magnitude. This has a sizeable effect on the derived SN rate since at
the magnitude of most of our SNe, the detection efficiency varies
rapidly with magnitude.  We estimated the size of the effect using the
current best estimate of $\pm 0.2$ mag uncertainty in the APM
calibration of our fields; this contributes $\pm 0.10 h^2$ SNu to the
rate uncertainty (in the sense that brighter assigned magnitudes
reduces the derived rate).

\paragraph*{Efficiency determination} Detection efficiencies were determined
using a Monte-Carlo simulation which was statistically limited (one
hundred fake SNe were added to each image). Also, models were used for
the distance of the SN to the host, and the host galaxy magnitude
distribution (assumed to be representative of the total galaxy
population). Figure 1d shows, however, that the detection efficiency
depends only weakly upon the magnitude difference between the host
galaxy and the SN and therefore upon the position of the SN on the
host and the host magnitude distribution.  We estimated less than 5\%
uncertainty on the efficiency from using these
assumptions. Altogether, uncertainties in the calculation of the
efficiency amount to $\pm 0.08 h^2$SNu uncertainty on the rate.

\paragraph{Range of Type Ia SN lightcurves} Control times were calculated using a single template lightcurve with a peak
brightness calibrated using the mean of the three observed SNe,
therefore making the assumption that Type Ia SNe are standard
candles. However, the observed rms scatter in peak brightness for Type
Ia SNe could be as big as 0.5 mag depending on the sample used (Riess
et al. 1995, Vaughan et al, in preparation). A correlation between
peak brightness and lightcurve width (Phillips 1993, Hamuy et
al. 1995, Riess et al. 1995) can nevertheless be used to reduce the
scatter to 0.21 mag or better (Hamuy et al. 1995 Riess et
al. 1995). We therefore estimate a one sigma systematic effect on the
measured rate assuming that the overall scatter in brightness of 0.5
comes from two independent sources : i) An ``intrinsic'' scatter of
0.21 mag, independent of lightcuve width. This was estimated by
altering the peak magnitude of the template lightcurve by
$0.21/\sqrt{3}$ mag which had the effect of changing the rate by $0.07
h^2$ SNu.  ii) A contribution of 0.45 $(=\sqrt{0.5^2-0.21^2})$ mag
correlated with lightcurve width. This was estimated by altering the
peak brightness by $0.45/\sqrt{3}$ mag and correspondingly the width
of the template lightcurve. To do this we used an approximation for
the width-magnitude relation, following the ``stretch factor'' method
of Perlmutter et al (1996) which reproduces the results of Hamuy et al
(1995) and Riess et al (1995). This changes the rate by $0.18 h^2$
SNu.  The overall uncertainty due to the intrinsic and calibratable
dispersion of Type Ia lightcurves therefore amounts to $\pm 0.19 h^2$
SNu on the rate. Note that this is a conservative estimate since
magnitude-limited samples give observed dispersions in peak magnitude of
$\sim 0.35$mag as compared to $\sim 0.5$mag for volume-limited
samples.

\paragraph*{Scanning efficiency} One SN candidate -- the faintest --
was not followed up (see section 2). If this was indeed a Type Ia
event, then the estimate of the rate increases by $0.27h^2$ SNu.

For any assumed galaxy counts, the main contribution to the systematic
uncertainty comes from the range of Type Ia SN lightcurves.
High-redshift supernovae from ongoing searches, including the recent
11 discoveries of this group, will soon bring down the statistical
uncertainty so that the systematic uncertainty will limit the accuracy
of high-redshift SNe rate measurements. The sensitivity to the assumed
galaxy counts was not included in this estimation and is discussed in
the next section. Assuming the Lilly et al counts for $N(z,R)$, we
estimate the total systematic error to be $^{+0.37}_{-0.25}$.

\section{Discussion}

\paragraph*{Galaxy counts}
 
To test the sensitivity of our result to the galaxy counts, 
we recalculated the rate using the model of
Gronwall \& Koo (1995) and that used by Glazebrook et al (1995).  

The galaxy counts of Gronwall \& Koo were kindly provided 
by C. Gronwall, and
are based on the analysis in Gronwall \& Koo (1995).  The model is a
``passive evolution'' model which has been constrained using galaxy
counts in various bands, principally $B_J$, and color and redshift
distributions for various ranges of $B_J$ (see Koo, Gronwall \&
Bruzual, 1993).  In determining their model from the data, $H_0=50
kms^{-1}Mpc^{-1}$ and $q_0=0.5$ were assumed. A non-standard local
luminosity function is assumed to minimize evolution required to fit
the counts.

The model used by Glazebrook et al (1995) was also used. This
model was derived using the local luminosity function of Loveday et al
(1992), the morphological mix given by Shanks et al (1984) and $K$
corrections based on spectral templates of Rocca-Volmerange \&
Guideroni (1988).  In determining the model, $q_0=0.5$ was assumed.  A
normalisation $\phi^{*}=0.03 (h/Mpc)^3$ was used in this analysis.

These models are quite different as can be seen on Figure~\ref{models} 
where a comparison of the $R$ magnitude
distribution between Lilly et al. counts and the counts derived from 
Gronwall \& Koo and Glazebrook et al. models show very substantial 
differences in the redshift range $0.2<z<0.6$, where our SN search is most 
sensitive. The rate we derive using the
model of Gronwall \& Koo is $1.61^{+1.05}_{-0.73} h^2$ SNu, almost a
factor of two higher than the value derived using the Lilly et al
counts. Using Glazebrook et al. model, we derive 
a value for the rate of
$1.27^{+0.83}_{-0.57} h^2$ SNu, which differs by 50\% from the
rate derived using the Lilly et al counts.

Before drawing any conclusion from these results, it should be noted that
unlike the galaxy counts derived from the above models, Lilly et al.
counts are based on data that are well-matched to our survey in magnitude and
redshift range, and only small amount of extrapolation was required in
converting from the $I$ to $R$ band. We therefore believe that the large 
differences
between the results reflects uncertainties in the extrapolation of the 
models of Gronwall \& Koo and Glazebrook et al. to match our data set and 
we did not quote any systematic uncertainty from galaxy counts in table 3. 
     
\paragraph*{Host galaxy inclination and extinction} The effect of host
galaxy inclination on detection efficiency and host galaxy luminosity
estimates should be taken into account when calculating supernovae
rates. Cappellaro et al. (1993b) and van den Bergh \& Tammann (1991)
have estimated the inclination correction factors for nearby searches.
In this analysis, both the search technique (in our case subtraction
of CCD images) and calculation of the galaxies' luminosities were done
differently than in most nearby searches, and the effects of galaxy
inclination should not be the same.

Galaxy inclination and extinction would reduce both the number of
supernovae detected and the galaxy visible luminosity. These effects
may therefore partially cancel in the calculation of the rate. A
complete analysis of this effect would require modeling of galaxy
opacities, which is beyond the scope of this paper. We therefore compare
our uncorrected value with uncorrected values for nearby searches, with
particular attention to CCD searches.

\paragraph*{Consistency check} This analysis is based on a subsample of 
data taken during winter 1994.  The larger sample of seven SNe was
discovered in approximately double this number of fields in three
different periods of data taking. Preliminary analysis of this data
show consistency with the results presented here.  As a further
consistency check we have examined the original data set for SNe which
are past maximum light. This was done by subtracting the search images
from the reference images (the reverse of the usual method) and
searching for positive signal as before. Two possible SNe were found
in this way, consistent with the number discovered on the rise.  Since
we have no further information on whether these are Type Ia SNe or
not, they have not been included in our determination of the rate.

\paragraph*{Detection efficiencies} 
The study of detection efficiencies as a function of SN magnitude is a
key element of this analysis. These detections efficiencies depend
upon many parameters and vary widely from field to field. It is
therefore essential to carefully and systematically estimate them.
The knowledge of these efficiencies will also be very useful for
estimating the effects of Malmquist bias on our sample of SNe. This
will be particularly important when using the distribution of measured
peak magnitudes to estimate $q_0$. At $z \sim 0.4$, the present mean
efficiency curve, applied to a Gaussian distribution of peak magnitude
with 0.2 magnitude intrinsic dispersion, would lead to a shift in the
derived value of $q_0$ of approximately $0.1$, if not taken into
account.

\paragraph*{Comparison with Other Measurement}
This is the first direct measurement of the Type Ia rate at high
redshift.  In their pioneering work, searching for high redshift
Supernovae, Hansen et al. (1989) discovered a probable Type II event
at $z = 0.28 $ and a Type Ia event at $z = 0.31$ (N\o rgaard-Nielsen
et al., 1989). No estimates of SN rates were published after the Type
Ia discovery, but beforehand they had concluded that their observation
was in mild disagreement with an expected number of Type Ia (based on
local rates) of $2.2-9.2$ (the range indicates the range of
determinations of the local rate from Van den Bergh et al, 1987 and
Cappellaro \& Turrato 1988).

Nearby supernovae rates have been carefully reanalyzed recently
(Cappellaro et al. 1993a \& 1993b, Turatto et al., 1994, Van den Bergh
and McClure, 1994, Muller et al. 1992) using more precise methods for 
calculating the
control times and correcting for inclination and over-exposure of the
nuclear regions of galaxies
in photographic searches.  The rate obtained for Type Ia SNe are
now consistent among these groups and vary between 0.2 $h^2$~SNu and
0.7 $h^2$~SNu depending on the galaxy types (E, Sa etc., higher rates
are found in later type galaxies).

In order to compare these rates with our measurement, one should
remember that (1) most local measurements have been based on
photographic data rather that CCD data as used here, (2) we did not
apply any correction for host galaxy absorption and inclination and
(3) at $z\sim$0.4 the ratios of galaxy type are different. Using
galaxy counts from Gronwall \& Koo in the range $0.35<z<0.45$ and
$21.75<R<22.25$ and their color classification of galaxies (Gronwall,
private communication), we estimate the relative fraction of galaxy
types in our sample to be 23\% E-S0, 15\%S0a-Sb and 62\% Sbc-Sd.
Combining this with the Type Ia rates measured by Cappellaro et al
(1993b) for E-S0, S0a-Sb and Sbc-Sd galaxy types, we can calculate the
local rate we should find if the mix of galaxies locally were the same
as the mix at $z\sim 0.4$. We obtain $0.53 \pm 0.25$ $h^2$ SNu. Our
measured value of $0.82 {^{+0.65}_{-0.45}}h^2$ SNu (where statistical
and systematic uncertainties have been combined), although slightly
higher, agrees with this value within the uncertainty and indicates
that Type Ia rates do not change dramatically out to $z\sim$0.4.
Note, however that correcting for host galaxy extinction and
inclination may change this conclusion.

Theoretical estimates of Type Ia SN rates have been derived from
stellar and galaxy evolution models. Calculations were done mostly for
elliptical galaxy type.  Earlier calculations predicted lower than
observed rates for Type Ia (Tornamb\`e \& Matteucci 1986, 1987,
Tornamb\`e 1989).  More recent calculations, based on evolutionary
models of elliptical galaxies, predict rates of $\sim 0.1 h^2$ SNu
(Ferrini \& Poggianti 1993). Assuming a factor of $\sim 2$ higher rate
in non-elliptical galaxies compared to ellipticals (Cappellaro et
al. 1993b) and a mix of galaxy types as above, we convert this to an
overall rate of Type Ia SNe at $z \sim 0.4$ in all galaxy types, and
derive a value of $\sim 0.37 h^2$ SNu.  Our total uncertainty of
$^{+0.65}_{-0.45}$ in the measurement presented in this paper does not
allow any firm conclusion but our observed rate seems to lie above
this theoretical prediction. There may be an increase of Type Ia rate
with redshift.  Ruiz-Lapuente, Burkert \& Canal (1995) predict
significant increase in rate for redshifts between 0.4 and 0.8
depending on the specific model they consider.  In the near future,
our ongoing high-$z$ SN search and others should provide enough data
to constrain the theoretical calculations.

\section*{Acknowledgments} This work was supported in part by the
National Science Foundation (ADT-88909616) and the U.S. Dept. of
Energy (DE-AC03-76SF000098).  We thank the La Palma staff \& observers
for carrying out service observations. We also thank Simon Lilly and 
Caryl Gronwall \&
David Koo for providing their galaxy counts prior to
publication, and Richard Kron for useful discussions. We are grateful
to the referee, Enrico Cappellaro, for helpful suggestions which led to 
the improvement of this paper. 
I.M. Hook acknowledges
a NATO fellowship. R. Pain thanks Gerard Fontaine of CNRS-IN2P3 and Bernard
Sadoulet of CfPA, Berkeley for encouraging his participation to this project.
\newpage
\input{tab1.tex}

\clearpage
\newpage
\input{tab2.tex}

\clearpage
\newpage
\input{tab3.tex}


\clearpage

\clearpage

\begin{figure}
\plotone{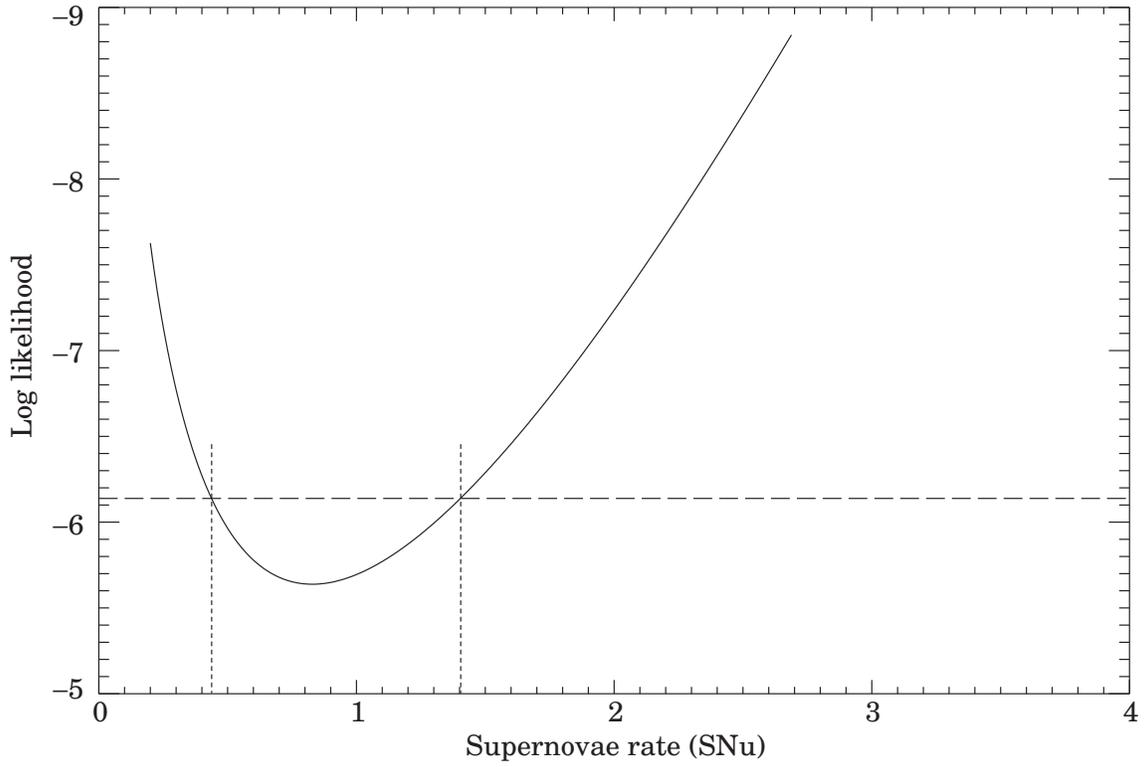}
\caption{Results from the maximum-likelihood fit of the observed
distribution of SN events to the expected distribution, $N_{exp}(z)$. The
dashed vertical lines show the $\pm 1 \sigma$ uncertainty in the result.
\label{maxl} }
\end{figure}

\begin{figure}
\plotone{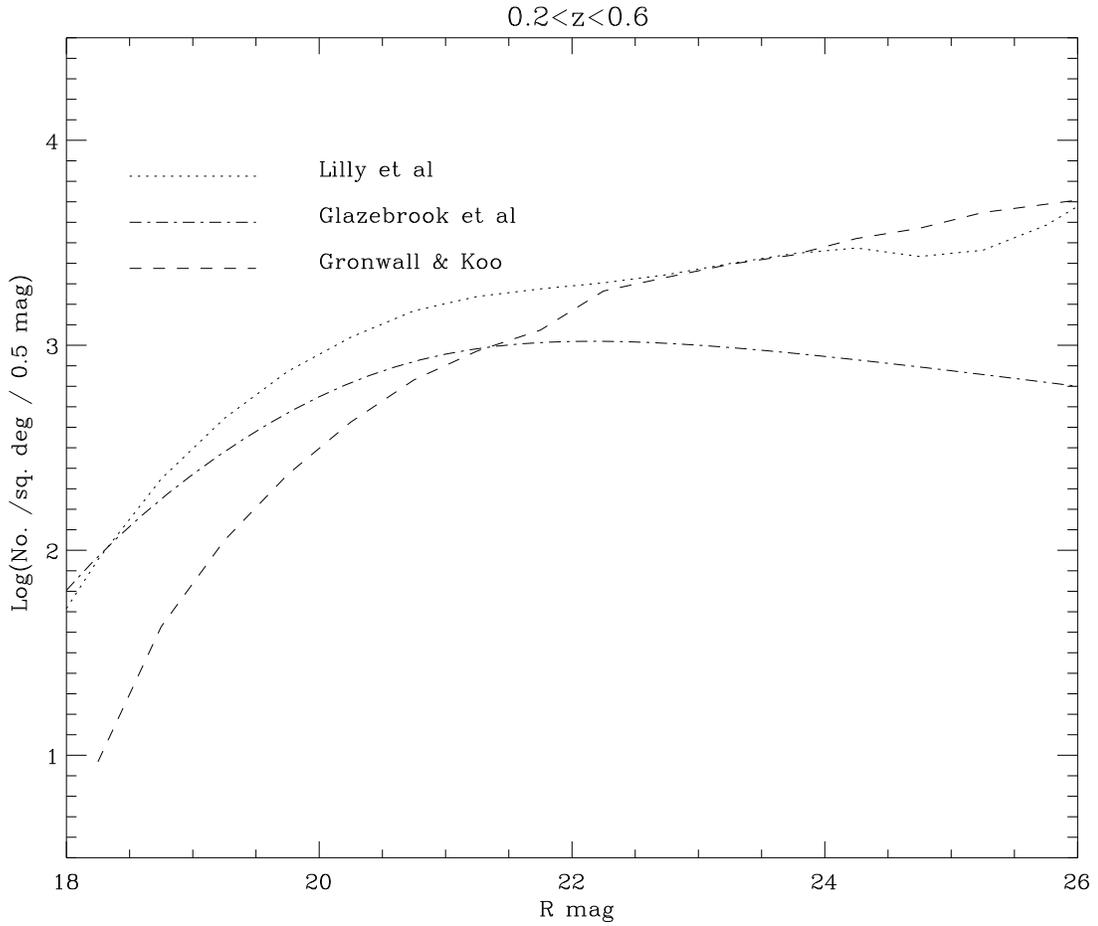}
\caption{Comparison of the galaxy N(R) in the redshift range
$0.2<z<0.6$.  The counts of Lilly et al (derived from the analysis of
Lilly et al 1995), and the models of Gronwall \& Koo (1995) and
Glazebrook et al (1995) are shown. The fluctuations in the curves
reflect the statistical fluctuations in the data from which the models
were derived. \label{models} } 
\end{figure}

\end{document}

%% file: tab1.tex
\begin{table} 
\begin{center}

\begin{tabular}{|lcrcc|}\hline
 Name & {\it z} & Host  & Dist. from    & Discovery  \cr
      &         & R mag & Host Core     & R mag \cr\hline
 1994H  & 0.375 & 21.1 & $1.0''$  & $ 21.9 $ \cr
 1994al & 0.420 & 21.2 & $1.0''$  & $ 22.6 $ \cr
 1994F  & 0.354 & 20.2 & $2.8''$  & $ 22.0 $ \cr\hline
\end{tabular}
\end{center}
\caption{Summary of the three SNe found in the survey
data described in the text.}  
\label{sntab}
\end{table}

%% file: tab2.tex
\begin{table} 
\begin{center} \begin{tabular}{|l|rrrrrr|}\hline $R
\backslash {\it z}$& 0.05 & 0.15 & 0.25 & 0.35 & 0.45 & 0.55 \cr\hline

 18.5 &   12.94 &   824.9  &   2808  &  3671 &   881.4 &  88.43 \cr
 19.5 &   7.942 &   322.6  &   2451  &  5368 &   3691  &  2031 \cr 
 20.5 &   3.799 &   154.7  &   943.1 &  3520 &   6050  &  6430 \cr 
 21.5 &   1.647 &   87.31  &   397.2 &  1281 &   3059  &  6699 \cr 
 22.5 &   1.211 &   30.60  &   226.4 &  587.2&   1243  &  2728 \cr 
 23.5 &   1.126 &   11.26  &   88.16 &  340.4&   744.2 &  1212 \cr
 24.5 &  0.1458 &   18.56  &   17.07 &  105.8&   351.5 &  799.2 \cr
 25.5 &   0.00  &   9.939  &   60.59 &  27.68&   65.53 &  280.5 \cr\hline
\end{tabular}
\end{center}
\caption{Total luminosity in units of $10^{10}\Lbsun$
given by the counts of Lilly et al (1995) and assuming $q_0=0.5$ 
, in the survey area of
1.73 sq deg.  Bin widths are 0.1 in $z$, 1.0 mag in R, centered on the
values shown.} 
\label{ntot}
\end{table}

%% file: tab3.tex
\begin{table}
\begin{center}
\begin{tabular}{|lc|}\hline
 Source & uncertainty \cr \hline
Luminosity estimate                     &   0.09 \cr
Cluster contribution                    &   0.02 \cr
Galaxy extinction                       &   0.01 \cr
APM calibration                         &   0.10 \cr
Detection efficiency                    &   0.08 \cr
Range of Ia lightcurves                 &   0.19 \cr
Scanning efficiency                     &$-0.00+0.27$ \cr\hline
 
Total syst. uncertainty                 &$-0.25+0.37$ \cr\hline

\end{tabular}
\end{center}
\caption{Systematic Uncertainties.  Uncertainties in the
rate are in $h^2$ SNu.  All uncertainties were estimated using the
Lilly et al counts for the magnitude-redshift distribution of
galaxies, N(R,z). Note that no estimate of systematic uncertainties  
from galaxy counts and from Galaxy inclination and extinction was made.}
\label{systab}
\end{table}

%% file: rates.bbl
\begin{thebibliography}{}

\bibitem{bh}
{Burstein, D., Heiles, C.}
\newblock{1982, \aj, 87, 1165}

\bibitem{cardelli}
{Cardelli, J. A., Clayton, G. C. Mathis, J. S}
\newblock{1989, \apj, 345, 245}

\bibitem{capturr}
{Cappellaro E., Turatto M.}
\newblock{1988, A\&A, 190, 10}

\bibitem{cap1993a}
{Cappellaro E., Turatto M., Benetti S., Tsvetkov, D. Yu, Bartunov, O. S.
\& Makarova I. J.
1993a, A\&A, 268, 472}

\bibitem{cap1993b}
{Cappellaro E., Turatto, M., Benetti, S., Tsvetkov, D. Yu, Bartunov, O. S.
\& Makarova I. J.
1993b, A\&A, 273, 383}

\bibitem{Ferrini}
{Ferrini F. \& Poggianti B.} 
\newblock{1993, \apj, 410, 44}

\bibitem{kgb}
{Glazebrook, K. Ellis, R. Santiago, B. Griffiths, R.}
\newblock {1995, MNRAS, 275, L19}

\bibitem{GK}
{Gronwall, C. \& Koo, D.}
\newblock {1995, \apjl, 440, L1}

\bibitem{gunnHoessel}
{Gunn, Hoessel, Oke, B} 
\newblock {1986, \apj, 306, 30}

\bibitem{hansen}
{Hansen, L., J\o rgensen, H. E.,  N\o rgaard-Nielsen, 
H. U., Ellis, R. S., Couch, W. J.
1989, A\&A, 211, L9}

\bibitem{Hamuy3}
{Hamuy, M., Phillips, M.~M., Maza, J., Suntzeff, N.B., Schommer, R.,
\& Aviles, R.
1995, \aj, 109, 1}

\bibitem{kim}
{Kim, A., Goobar,. A, Perlmutter, S.}
\newblock {1996, \pasp, 108, 190}

\bibitem{koo}
{Koo, D., Gronwall, C., Bruzual, A. G.}
\newblock {1993, \apj, 415, L21}

\bibitem{Leib}
{Landolt, A. U.}
\newblock {1992, \aj, 104, 340}

\bibitem{Leib}
{Leibundgut, B.}
\newblock {1988, Ph. D Thesis, University of Basel}

\bibitem{Lilly}
{Lilly, S., Tresse, L., Hammer, F., Cramptopn, D., Le Fevre, O.}
\newblock {1995, \apj, 455, 108.}

\bibitem{Loveday}
{Loveday, J. Peterson, B. A., Efstathiou, G. Maddox, S. J.}
\newblock {1992, \apj, 400, L43}

\bibitem{muller}
{Muller et al.}
\newblock {1992, \apjl, 384, L9}

\bibitem{possy} 
{McMahon R. G. and Irwin M. J. 1992 , Digitised Optical Sky Surveys,}
\newblock {eds H. T. MacGillivray and E. B. Thomson, Kluwer p 417.}

\bibitem{Nielsen}
{N\o rgaard-Nielsen, H. U. et al., 1989, Nature, Vol 339, 523}

\bibitem{sne1994}
{Perlmutter, S., et al.}
\newblock {1994. IAUC no. 5956}

\bibitem{sn1992bi}
{Perlmutter, S., et al.}
\newblock {1995a. \apjl, 440, L41}

\bibitem{teleg2}
{Perlmutter, S. et al}
\newblock {1995b. IAUC no. 6263}

\bibitem{teleg3}
{Perlmutter, S. et al}
\newblock {1995c. IAUC no. 6270}

\bibitem{stretch}
{Perlmutter, S. et al. 1996 in
Proceedings of NATO Advanced Study Institute on Thermonuclear
Supernovae, R. Canal, P. Ruiz-Lapuente and J. Isern, Eds., (Kluwer
Dordrecht), in press}

\bibitem{phillips}
{Phillips, M.M.}
\newblock {1993, \apjl, 413, L105}

\bibitem{riess}
{Riess, A. G., Press, W. H., \& Kirshner, R. P. }
\newblock {1995, \apjl, 438, L17}

\bibitem{rv}
{Rocca-Volmerange, B., Guideroni, B.}
\newblock {1988, A\&AS, 75, 93}

\bibitem{ruiz}
{Ruiz-Lapuente, P. Burket, A., Canal, R.}
\newblock {1995, \apj, 447, L69}

\bibitem{shanks}
{Shanks, T. S,  Stevenson, P. R., Fong, R., MacGillivray, H. T}
\newblock {1984, MNRAS, 206, 767}

\bibitem{tornambe86}
{Tornamb\`e A. \& Matteucci F.}
\newblock{1986, MNRAS, 223, 69}

\bibitem{tornambe87}
{Tornamb\`e A. \& Matteucci F.}
\newblock{1987, \apj, 318, L25}

\bibitem{tornambe89}
{Tornamb\`e}
\newblock{1989, MNRAS, 239, 771}

\bibitem{turrato}
{Turatto M., Cappellaro \& E., Benetti S.
\newblock{1994,  A. J.,  108}, 202.}

\bibitem{vdbmc}
{Van den Bergh S. \& McClure R. D.}
\newblock{1994, \apj, 425, 205}

\bibitem{vdb}
{Van Den Bergh, S. McClure, R. D., Evans, R.}
\newblock {1987, \apj, 323, 44}

\bibitem{vdbtam}
{Van Den Bergh, S., Tamman, G. A.}
\newblock {1991, \araa, 29, 363}

\end{thebibliography}
